# OCMR (v1.0)—Open-Access Multi-Coil k-Space Dataset for Cardiovascular Magnetic Resonance Imaging

Chong Chen,[1] Yingmin Liu,[2] Philip Schniter,[3] Matthew Tong,[4] Karolina Zareba,[4] Orlando Simonetti,[2,4] Lee Potter,[2,3] and *Rizwan Ahmad[1,2,3]

**1** Biomedical Engineering, The Ohio State University, Columbus OH, USA

**2** Davis Heart and Lung Research Institute, The Ohio State University, Columbu OH, USA

**3** Electrical and Computer Engineering, The Ohio State University, Columbus OH, USA

**4** Internal Medicine, The Ohio State University, Columbus OH, USA

**\*** Corresponding author:

| | |
|---:|:---|
| **Name** | Rizwan Ahmad |
| **Department** | Biomedical Engineering |
| **Institute** | The Ohio State University |
| **Address** | 460 W 12th Ave, Room 318 |
| | Columbus OH 43210, USA |
| **E-mail** | ahmad.46@osu.edu |

**Abstract**

Cardiovascular MRI (CMR) is a non-invasive imaging modality that provides excellent soft-tissue contrast without the use of ionizing radiation. Physiological motions and limited speed of MRI data acquisition necessitate development of accelerated methods, which typically rely on undersampling. Recovering diagnostic quality CMR images from highly undersampled data has been an active area of research. Recently, several data acquisition and processing methods have been proposed to accelerate CMR. The availability of data to objectively evaluate and compare different reconstruction methods could expedite innovation and promote clinical translation of these methods. In this work, we introduce an open-access dataset, called OCMR, that provides multi-coil k-space data from 53 fully sampled and 212 prospectively undersampled cardiac cine series.

# Introduction

Cardiovascular Magnetic Resonance (CMR) is an MRI-based non-invasive imaging tool that provides a comprehensive assessment of the cardiovascular system. Although there are competing modalities for nearly every CMR application, there is no single imaging modality that can match CMR's versatility and accuracy. A single CMR exam can accurately evaluate the cardiovascular structure, function, morphology, perfusion, viability, and hemodynamics (1). Despite the growing evidence of its advantages over other modalities and its potential as a "one-stop-shop" diagnostic tool, the role of CMR in clinical cardiology remains limited. One major impediment to broader usage of CMR is inefficient acquisition, which makes traditional CMR exams excessively long (often longer than an hour) and diminishes their efficiency and cost-effectiveness relative to other modalities. Also, the growing demand for free-breathing imaging and the emergence of high dimensional imaging, e.g., 5D whole-heart imaging (2), demand acceleration rates that cannot be realized with existing technology. Therefore, there is a growing need to develop new techniques for accelerated CMR.

CMR data generated by clinical MRI scanners have at least one coil and two spatial dimensions. Many applications bring additional dimensions, e.g., time, leading to a high-dimensional array. Although these multi-dimensional images contain a huge number of pixels, they are highly structured, and this structure can be exploited to recover them from relatively few measurements. Several advanced data acquisition and processing techniques have been proposed and validated for a diversity of CMR applications. Yet, the clinical translation of these innovations has been slow. The lack of a platform to objectively and fairly compare different techniques has been a major impediment to rapid clinical translation. Moreover, for machine-learning-based methods, which have become increasingly popular for MRI reconstruction (3), the training data act as an implicit part of the method. With this motivation, we launch an open-access dataset for raw k-space CMR data that can be used to evaluate and compare different Cartesian sampling and reconstruction techniques.

# CMR Applications

For a comprehensive functional and anatomic evaluation of the cardiovascular system, a CMR exam consists of several individual scans. Cardiac cine provides a movie of the beating heart and is the most common CMR application (4). Typically, cine movies are sequentially collected from a stack of 2D slices in the short-axis orientation (see Figure 1), covering the entire heart. Slices in the long-axis orientation are routinely acquired to visualize wall-motion abnormalities and valvular apparatus. The images from the cine stack are analyzed to quantitatively and qualitatively assess the heart pumping function. Phase-contrast (PC)-based flow imaging is another common CMR application that allows one to measure blood flow through the cardiac valves, inside the cardiac chambers, and in the vessels (5). Other CMR applications routinely used in clinical practice include first-pass perfusion (FPP) (6), late gadolinium enhancement (LGE) (7), cardiac parametric mapping (8), and Magnetic Resonance Angiography (MRA) (9). In FPP, the dynamics of a Gadolinium-Based Contrast Agent (GBCA), as it perfuses through the myocardium, are captured at rest or under vasodilator or inotropic stress. FPP is used to detect perfusion defects in ischemic heart disease and intracardiac shunts in congenital heart disease. LGE images signal enhancement from the tissue accumulation of GBCA and is used to detect myocardial scarring or fibrosis at the cellular level. Cardiac parametric mapping is a quantitative technique commonly used to characterize the myocardium. It can provide spatially resolved maps of the tissue relaxation times: T1, T2, or T2*. Abnormal relaxation times are suggestive of tissue abnormalities, including myocardial inflammation, edema, diffuse interstitial fibrosis or myocardial infiltration. Finally, MRA is used to visualize blood vessels and identify vascular abnormalities. Other less common but emerging applications of CMR include strain imaging, stress T1 mapping (10), elastography (11), displacement encoding imaging (12), and diffusion tensor imaging (13).

In terms of methodology, there are several ways to partition CMR imaging. For example, CMR acquisition can be performed in planar (2D) or volumetric (3D) mode. Although 3D acquisition is becoming popular, 2D acquisition remains more prevalent in clinical setups. Also, some CMR applications (e.g., cine, flow, FPP, and MRA) image both spatial and temporal dimensions, while others (e.g., LGE and parametric mapping) image only spatial dimensions. The data for CMR can be collected in real-time or single-shot mode (14) or in segmented mode (4). A segmented acquisition aggregates measurements from different heartbeats and thus relies on breath-holding and regular heart rhythm. In contrast, a real-time or single-shot acquisition does not mix measurements from different heartbeats, obviating the need for breath-holding or regular heart rhythm. This strategy, however, requires higher acceleration rates, leading to degraded image quality. For contrast-enhanced dynamic applications, the data is almost always collected in real-time or single-shot mode, often under free-breathing conditions. For other applications, breath-held segmented CMR remains the standard acquisition mode, with real-time or single-shot acquisition reserved for patients who cannot hold their breath or who are arrhythmic. For applications where breath-holding is not an option and real-time acquisition would lead to unreasonably high acceleration rates, respiratory gating can be employed (15). The gating is typically performed prospectively, using navigator echoes. More recently, techniques based on free-running acquisition have become popular, at least in research settings. In these techniques, the measurements are continuously acquired and then retrospectively sorted into respiratory and cardiac bins using self-gating (16). Finally, most clinical studies are performed with Cartesian sampling of k-space, but radial sampling has become increasingly common (17).

## Data Modeling

In MRI, the measured k-space data are samples of the 2D or 3D continuous Fourier transform of the underlying 2D or 3D image. Many CMR applications include additional encoding dimensions, e.g., the time dimension is included for dynamic applications and the velocity dimension is included for phase-contrast CMR. On all modern MRI scanners, the CMR data are collected in parallel, using multiple receive coils, which are distributed around the anatomy of interest; this mode of acquisition is known as parallel MRI. For a dynamic CMR application, the k-space signal measured from the $c^{\text{th}}$ coil at the $t^{\text{th}}$ time frame can be expressed as (18)

$$\boldsymbol{y}_c^{(t)} = \boldsymbol{D}^{(t)} \boldsymbol{F} \boldsymbol{S}_c \boldsymbol{x}^{(t)} + \boldsymbol{w}_c^{(t)}, \qquad [1]$$

where $\boldsymbol{x}^{(t)} \in \mathbb{C}^N$ is the vectorized 2D or 3D image at time frame $t$, $\boldsymbol{S}_c \in \mathbb{C}^{N \times N}$ is a diagonal matrix containing the sensitivity map of the $c^{\text{th}}$ coil, $\boldsymbol{F} \in \mathbb{C}^{N \times N}$ is the 2D or 3D discrete Fourier transform matrix, $\boldsymbol{D}^{(t)} \in \mathbb{R}^{M \times N}$ is a binary, diagonal undersampling matrix, and $\boldsymbol{w}_c^{(t)} \in \mathbb{C}^M$ is additive white Gaussian noise. Note that the sampling pattern can change across time but stays the same across coils, while the sensitivity maps change across coils but not across time.

The sampling trajectories through the k-space are determined by the magnetic field gradient waveforms and can be Cartesian or non-Cartesian in nature. A Cartesian trajectory represents parallel lines through k-space, with one fully sampled dimension, known as "frequency encoding." The other one or two dimensions, known as "phase encodings," are often undersampled to reduce the acquisition time. Typically, one k-space line, called the "readout," is collected after the application of each radio-frequency (RF) pulse, and this process is repeated several times to adequately sample the k-space. The ratio $R = N/M$ is often referred to as the acceleration rate. In commonly used non-Cartesian trajectories, each readout traverses a radial or spiral path. Compared to Cartesian sampling, non-Cartesian sampling can provide more efficient coverage of k-space and can yield an "incoherent" forward operator that is more conducive to compressed sensing (CS) reconstruction. However, Cartesian sampling, due to its higher tolerance to system imperfections and a more extensive record of success, remains the method of choice in clinical practice.

## Image Reconstruction

For CMR, image reconstruction can be performed in k-space or in the image domain. The most common k-space method is GRAPPA (19). In GRAPPA, interpolation weights, called kernels, which express an unacquired k-space sample as a weighted average of the acquired k-space samples across all coils, are learned from a fully sampled region of k-space. These kernels are then used to estimate unacquired samples from acquired ones. The estimated k-space from each coil is subsequently converted to an image by applying the inverse 2D or 3D Fourier transform. Finally, the images from individual coils are combined to generate a "coil-combined" image. SPIRiT (20) is a k-space method that generalizes GRAPPA by expressing each k-space sample as a weighted combination of its neighboring samples (acquired or unacquired) from all coils. The unacquired samples are then inferred by iteratively enforcing this self-consistency across the k-space. SPIRiT has been combined with sparsity-based priors to yield reconstruction from highly undersampled data (21). In addition to GRAPPA and SPIRiT, calibrationless k-space methods have also been proposed, where the linear dependency among coils and the neighboring k-space samples is exploited via rank deficiency of the lifted block-Hankel matrices (22). Although these methods offer the potential of high acceleration rates, their clinical utility is currently limited by computation efficiency.

In contrast to GRAPPA and SPIRiT, SENSE-based methods attempt to infer the image series, $\boldsymbol{x}^{(t)}$, from Eq. 1 (23). The sensitivity map, $\boldsymbol{S}_c$, changes with the orientation of the imaging plane as well as from patient to patient. Therefore, both $\boldsymbol{S}_c$ and $\boldsymbol{x}^{(t)}$ are unknowns, leading to a bilinear recovery problem. Although several methods have been proposed to iteratively solve for both $\boldsymbol{S}_c$ and $\boldsymbol{x}^{(t)}$ (24, 25), a more common practice is to linearize the problem by estimating $\boldsymbol{S}_c$ first, using techniques such as ESPIRiT (26) or the one proposed by Walsh et al. (27). For static applications, $\boldsymbol{S}_c$ is estimated from low-resolution, fully sampled data, which may be available from the measured data or from separately acquired calibration data. For dynamic applications, $\boldsymbol{S}_c$ can be estimated from the time-averaged k-space samples, provided such averaging leads to a fully sampled region in k-space. Once $\boldsymbol{S}_c$ is estimated, the data measured from all $C$ receiver coils can be collectively expressed as

$$\underbrace{\begin{bmatrix} \boldsymbol{y}_1^{(t)} \\ \boldsymbol{y}_2^{(t)} \\ \vdots \\ \boldsymbol{y}_C^{(t)} \end{bmatrix}}_{\boldsymbol{y}^{(t)} \in \mathbb{C}^{MC \times 1}} = \underbrace{\begin{bmatrix} \boldsymbol{P}^{(t)} \boldsymbol{F} \boldsymbol{S}_1 \\ \boldsymbol{P}^{(t)} \boldsymbol{F} \boldsymbol{S}_2 \\ \vdots \\ \boldsymbol{P}^{(t)} \boldsymbol{F} \boldsymbol{S}_C \end{bmatrix}}_{\boldsymbol{A}^{(t)} \in \mathbb{C}^{MK \times N}} \boldsymbol{x}^{(t)} + \underbrace{\begin{bmatrix} \boldsymbol{w}_1^{(t)} \\ \boldsymbol{w}_2^{(t)} \\ \vdots \\ \boldsymbol{w}_C^{(t)} \end{bmatrix}}_{\boldsymbol{w}^{(t)} \in \mathbb{C}^{MC \times 1}}. \quad [2]$$

The signal model in Eq. 2 represents one frame in a dynamic image sequence over $t$. The signal model for the entire sequence of $T$ frames can be expressed by stacking the vectors $\boldsymbol{y}^{(t)}$, $\boldsymbol{x}^{(t)}$, and $\boldsymbol{w}^{(t)}$ across $t$ into $\boldsymbol{y}$, $\boldsymbol{x}$, and $\boldsymbol{w}$, respectively, and embedding the matrices $\boldsymbol{A}^{(t)}$ into the larger block diagonal matrix $\boldsymbol{A}$, i.e.,

$$\underbrace{\begin{bmatrix} \boldsymbol{y}^{(1)} \\ \boldsymbol{y}^{(2)} \\ \vdots \\ \boldsymbol{y}^{(T)} \end{bmatrix}}_{\boldsymbol{y} \in \mathbb{C}^{MCT \times 1}} = \underbrace{\begin{bmatrix} \boldsymbol{A}^{(1)} & 0 & 0 & \cdots & 0 \\ 0 & \boldsymbol{A}^{(2)} & 0 & \cdots & 0 \\ \vdots & & \ddots & & \vdots \\ 0 & 0 & 0 & \cdots & \boldsymbol{A}^{(T)} \end{bmatrix}}_{\boldsymbol{A} \in \mathbb{C}^{MCT \times NT}} \underbrace{\begin{bmatrix} \boldsymbol{x}^{(1)} \\ \boldsymbol{x}^{(2)} \\ \vdots \\ \boldsymbol{x}^{(T)} \end{bmatrix}}_{\boldsymbol{x} \in \mathbb{C}^{NT \times 1}} + \underbrace{\begin{bmatrix} \boldsymbol{w}^{(1)} \\ \boldsymbol{w}^{(2)} \\ \vdots \\ \boldsymbol{w}^{(T)} \end{bmatrix}}_{\boldsymbol{w} \in \mathbb{C}^{MCT \times 1}} \quad [3]$$

$$\boldsymbol{y} = \boldsymbol{A}\boldsymbol{x} + \boldsymbol{w}. \quad [4]$$

As a consequence of the AWGN model on $\boldsymbol{w}$, maximum likelihood (ML) estimation of $\boldsymbol{x}$ from $\boldsymbol{y}$ reduces to least-squares fitting, i.e., $\arg\min_{\boldsymbol{x}} \|\boldsymbol{y} - \boldsymbol{A}\boldsymbol{x}\|_2^2$. ML gives good performance if $MCT \geq NT$ and $\boldsymbol{A}$ is well-conditioned. With multiple receive coils ($C > 1$), these conditions can be met at low to moderate acceleration rates. However, this is rarely the case under moderate to high acceleration (i.e., $R \geq 3$). With significant acceleration, it is critically important to exploit prior knowledge of image structure.

The traditional approach to exploiting such prior knowledge is to formulate and solve a "variational" optimization problem of the form (28)

$$\widehat{\boldsymbol{x}} = \arg\min_{\boldsymbol{x}} \left\{ \|\boldsymbol{y} - \boldsymbol{A}\boldsymbol{x}\|_2^2 + \phi(\boldsymbol{x}) \right\}, \quad [5]$$

where the regularization term $\phi(\cdot)$ encodes prior knowledge about the image. It is common to use a regularization of the form $\phi(\boldsymbol{x}) = \lambda \|\boldsymbol{\Psi}\boldsymbol{x}\|_1$, where $\boldsymbol{\Psi}$ is a known transform (such as the finite-difference operator, for total-variation regularization), $\lambda > 0$ is a tunable weight, and the $\ell_1$ norm rewards sparsity in the transform outputs, $\boldsymbol{\Psi}\boldsymbol{x}$. This is sometimes referred to as "analysis" based

CS. This form of recovery has been included in the default image-processing frameworks offered by several MRI vendors. Due to the rich structure in images and image sequences, the use of composite regularization, i.e., $\phi(\boldsymbol{x}) = \sum_{l=1}^{L} \lambda_l \|\boldsymbol{\Psi}_l \boldsymbol{x}\|_1$ with $L > 1$, has been shown to improve image recovery quality (29). When $L > 1$, the tuning of $\{\lambda_l\}$ is nontrivial and becomes an important determinant of recovery performance (30).

More recently, machine-learning-based techniques have been shown to outperform CS recovery methods. Some of these techniques aim to learn the entire end-to-end mapping from the undersampled k-space measurements $\boldsymbol{y}$ (or the aliased image $\boldsymbol{A}^\mathsf{H}\boldsymbol{y}$) to the recovered image $\boldsymbol{x}$ (e.g., (31, 32)). Considering that the forward model $\boldsymbol{A}$ (and thus the aliasing pattern) changes from one scan to the next, even within a single CMR exam, such methods must be trained over a large and diverse data corpus or limited to a specific application. Other methods train a *scan-specific* convolutional neural network (CNN) on a fully-sampled region of k-space and then use it to interpolate missing k-space samples (33). For training, these methods do not require separate training data but rather many examples of small, fully sampled patches in k-space. Due to the large number of trainable parameters in a CNN, however, such methods require more patches than are typically acquired in GRAPPA or SPIRiT-type acquisitions, which limits the acceleration that can be achieved. So-called "unrolled" methods are inspired by classic variational optimization methods and alternate between processing stages that enforce data-consistency (exploiting knowledge of $\boldsymbol{A}$) and trainable CNN stages that enforce image structure (34, 35). Such methods require a large number of fully sampled k-space data scans for training. Also, since CNN training occurs in the presence of scan-specific forward models, generalization from training to test scenarios remains an open question (36). Plug-and-play (PnP) methods (37) provide another avenue to perform MRI recovery (38). Such methods can combine machine-learning-based MRI-specific image denoisers with physics-driven data models. Other learning-based methods have been proposed based on bi-level optimization (e.g., (39)), adversarially learned priors (e.g., (40)), and autoregressive priors (e.g., (41)).

Compared to static MRI, the extension of learning-based techniques to CMR has been challenging, in part, due to the limited availability of training data for CMR. For some applications, like FPP, it is not feasible to collect fully sampled (i.e., $R = 1$) data due to the limited time-window defined by constrast dynamics. For other applications, however, it is feasible. For example, it is possible to collect fully sampled cardiac cine and flow data under long breath-holds.

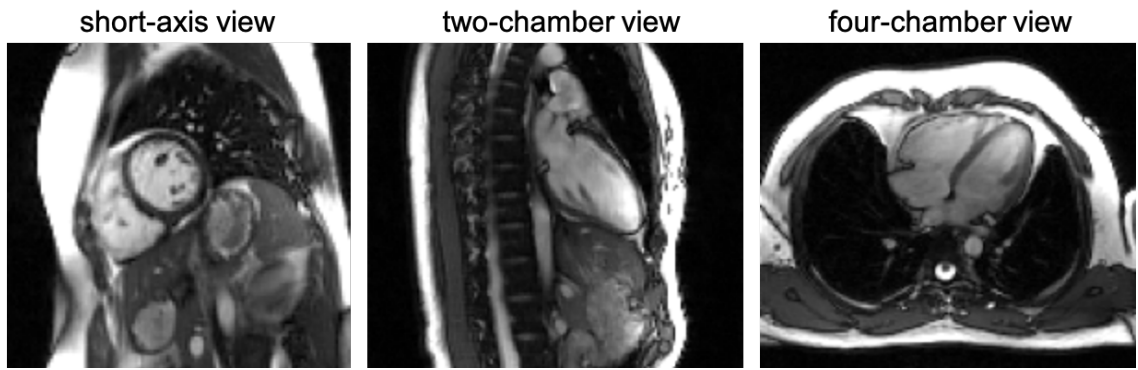

Figure 1: Real-time cardiac cine in three different views. Left: short-axis (SAX) view. Middle: two-chamber view. Right: four-chamber view. Both two-chamber and four-chamber views belong to the long-axis (LAX) view.

# OCMR Dataset

In this section, we describe the OCMR dataset and how to load it using Matlab and Python. The OCMR dataset is comprised of HDF5 files, with data in each file following the ISMRMRD format. There are a total of 265 data files, each representing a separate scan. Of them, 53 scans (comprising 81 slices) have no undersampling in the phase encoding direction, while the remaining 212 scans (comprising 842 slices) are prospectively undersampled. The datasets were collected on three Siemens MAGNETOM scanners: Prisma (3T), Avanto (1.5T), and Sola (1.5T). Each scan is assigned eight attributes, which allows one to download only a subset of the OCMR dataset. The attributes and their descriptions are listed in Table 1. More details about the dataset can here found here: `www.ocmr.info`—a dedicated website for the OCMR dataset.

## Reading Data using Matlab

Step 1: Download tar archive (ocmr_cine.tar.gz) by visiting `https://ocmr.s3.amazonaws.com/data/ocmr_cine.tar.gz`. This single file contains the entire dataset.

Step 2: Download Matlab wrapper read_ocmr.m and an example Matlab script example_ocmr.m from `https://github.com/MRIOSU/OCMR/tree/master/Matlab`.

Step 3: Download ISMRMRD libraries from `https://github.com/ismrmrd/ismrmrd/tree/master/matlab/%2Bismrmrd`. Note, only the subfolder '/+ismrmrd' is required.

| Attribute | Description |
|---|---|
| scn | This attribute identifies the field strength and type of scanner. The value of '15avan' selects datasets collected on 1.5T Siemens MAGNETOM Avanto; the value of '15sola' selects datasets collected on 1.5T Siemens MAGNETOM Sola; the value of '30pris' selects datasets collected on 3T Siemens MAGNETOM Prisma; and the value of 'all' selects datasets from all scanners. |
| smp | This attribute identifies different sampling patterns. The value of 'fs' selects fully sampled datasets; the value of 'pse' selects prospectively undersampled datasets with pseudo-random sampling; the value of 'uni' selects prospectively undersampled datasets with uniform undersampling. The value of 'all' selects both fully sampled and undersampled datasets. |
| ech | This attribute identifies asymmetric readout or echo. The value of 'asy' selects datasets with asymmetric echo, while the value of 'sym' selects datasets with symmetric echo. The value of 'all' selects data with and without asymmetric echo. See Figure 2. |
| dur | This attribute only applies to prospectively undersampled data and distinguishes long from short scans. The value of 'lng' selects datasets where the time dimension is at least 5 s long, while the value of 'shr' selects datasets where the time dimension is shorter than 5 s. The value of 'all' selects datasets regardless of the duration. All fully sampled datasets belong to 'shr'. |
| viw | This attribute defines the view of the imaging slice. The value of 'sax' selects datasets collected in the short-axis view, while the value of 'lax' selects datasets collected in the long-axis view. The value of 'all' selects datasets regardless of the view. See Figure 1. |
| sli | This attribute distinguishes individual slices from stacks. The value of 'ind' selects datasets collected as individual slices, while the value of 'stk' selects short-axis/long-axis stacks. The value of 'all' selects both individual slices and stacks. |
| fov | This attribute defines the presence of spatial aliasing, i.e., when the field-of-view (FOV) is smaller than the spatial extent of the content. The value of 'ali' selects datasets with spatial aliasing, while the value of 'noa' selects datasets without aliasing. The value of 'all' selects datasets regardless of spatial aliasing. Note, for SENSE-based methods, datasets with aliasing would require the utilization of multiple sets of sensitivity maps for artifact-free reconstruction (26). |
| sub | This attribute distinguishes patients from healthy volunteers. The value of 'pat' selects data collected from patients, while the value of 'vol' selects data collected from healthy volunteers. The value of 'all' selects datasets from both patients and healthy volunteers. |

Table 1: Attributes assigned to each OCMR scan.

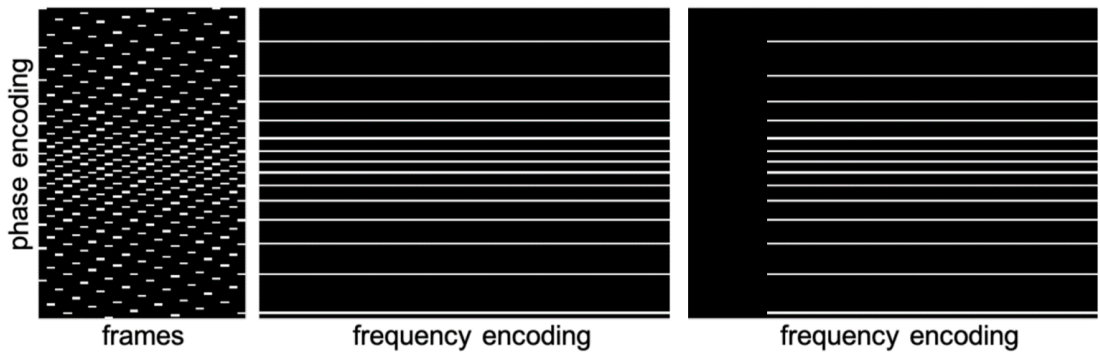

Figure 2: A representative pseudo-random sampling pattern (smp=pse). Left: phase encoding and frame (i.e., time) dimensions shown for one frequency encoding location. Middle: phase encoding and frequency encoding dimensions shown for the last frame. Right: same as the middle column but with asymmetric echo (ech=asy).

Step 4: Place read_ocmr.m, example_ocmr.m, and the subfolder '/+ismrmrd' in the same folder. Execute example_ocmr.m in Matlab and select the data file to be read; it will generate a nine-dimensional array (kData) for the k-space data and a structure (param) that captures acquisition parameters.

Note, the file listing the attributes of each data file can be found here: https://ocmr.s3.amazonaws.com/ocmr_data_attributes.csv.

## Reading Data using Python

Step 1: Download Python wrapper read_ocmr.py and example_ocmr.ipynb from at https://github.com/MRIOSU/OCMR/tree/master/Python.

Step 2: Execute example_ocmr.ipynb; it will generate a nine-dimensional array (kData) for the k-space data and a structure (param) that captures acquisition parameters. Note, this step eliminates the need to explicitly download the large tar file, ocmr_cine.tar.gz.

Step 3: By adjusting the filters in example_ocmr.ipynb, one could download selected files based on the attributes described in Table 1.

Note, the file listing the attributes of each data file can be found here: https://ocmr.s3.amazonaws.com/ocmr_data_attributes.csv.

## Data Structure

Once an HDF5 file is read into Matlab (or Python), it yields a k-space array, kData, and a structure, param. The array kData has nine dimensions: [kx, ky, kz, coil, phase, set, slice, rep, avg], which represent frequency encoding, first phase encoding, second phase encoding, coil, phase (time), set (velocity encoding), slice, repetition, and the number of averages, respectively. For example, an array with frequency encoding size 160, phase encoding size 120, number of coils 18, number of frames 60, number of slices 10 will have kData with these dimension: $160 \times 120 \times 1 \times 18 \times 60 \times 1 \times 10 \times 1 \times 1$. The second output, param, provides pertinent acquisition parameters. For example, param.FOV, param.TRes, param.flipAngle_deg, param.sequence_type specify FOV, temporal resolution, flip angle, and the type of sequence, respectively. The spatial resolution, $[\Delta x, \Delta y, \Delta z]$, which is a important image quality parameter, can be calculated by [param.FOV(1)/size(kData,1), param.FOV(2)/size(kData,2), param.FOV(3)/size(kData,3)]. Note, the readout dimension includes a factor of two oversampling, leading to param.FOV(1) that is twice the value selected on the scanner. Although not required, a common pre-processing step in most reconstruction methods is to downsample the readout dimension. This is typically achieved by taking 1D FFT along the readout dimension and retaining the central 50% of the FOV. The cropped data is then converted back to k-space by taking 1D inverse FFT along the readout.

## Anonymization

All scans included in OCMR have been de-identified, where Protected Health Information (PHI) as well as scan date and location have been removed. All datasets have been manually inspected to ensure that identifying facial features are not included.

# Conclusion and Future Directions

OCMR is the first open-access dataset that provides multi-coil k-space data for cardiac cine. The fully sampled scans are intended for quantitative comparison and evaluation of image reconstruction methods. The free-breathing, prospectively undersampled scans are intended to qualitatively evaluate the performance and generalizability of the reconstruction methods. The current version (v1.0) of OCMR includes 53 fully sampled scans and 212 prospectively undersampled scans for cardiac cine. In future iterations, we will expand the size of cardiac cine data and add data from other

CMR applications, including 2D phase-contrast MRI.

# Acknowledgments

This work was funded by the National Institutes of Health under grant NIH R01HL135489. We are thankful to John Wear from Engineering Technology Services for technical support.